\documentclass[aps,prb,floatfix,twocolumn,superscriptaddress]{revtex4-2}

\usepackage{mathrsfs}
\usepackage{graphicx}
\usepackage[english]{babel}
\usepackage{amsmath}
\usepackage{amssymb}

\usepackage{relsize}
\usepackage{CJK}
\usepackage{dsfont}
\usepackage{bm}

\newcommand{\me}{\mathrm{e}}
\newcommand{\mi}{\mathrm{i}}

\newcommand{\dif}{\mathrm{d}}

\allowdisplaybreaks
\begin{document}

\title{Geometric phases of mixed quantum states: A comparative study of interferometric and Uhlmann phases}

\author{Xu-Yang Hou}
\affiliation{School of Physics, Southeast University, Jiulonghu Campus, Nanjing 211189, China}

\author{Xin Wang}
\affiliation{School of Physics, Southeast University, Jiulonghu Campus, Nanjing 211189, China}
\author{Zheng Zhou}
\affiliation{School of Physics, Southeast University, Jiulonghu Campus, Nanjing 211189, China}
\author{Hao Guo}
\email{guohao.ph@seu.edu.cn}
\affiliation{School of Physics, Southeast University, Jiulonghu Campus, Nanjing 211189, China}
\author{Chih-Chun Chien}
\email{cchien5@ucmerced.edu}
\affiliation{Department of physics, University of California, Merced, CA 95343, USA}
\begin{abstract}
Two geometric phases of mixed quantum states, known as the interferometric phase and Uhlmann phase, are generalizations of the Berry phase of pure states. After reviewing the two geometric phases and examining their parallel-transport conditions, we specify a class of cyclic processes that are compatible with both conditions and therefore accumulate both phases through their definitions, respectively. Those processes then facilitate a fair comparison between the two phases. We present exact solutions of two-level and three-level systems to contrast the two phases. While the interferometric phase exhibits finite-temperature transitions only in the three-level system but not the two-level system, the Uhlmann phase shows finite-temperature transitions in both cases. Thus, using the two geometric phases as finite-temperature topological indicators demonstrates the rich physics of topology of mixed states.
\end{abstract}

\maketitle
\section{Introduction}
Geometric phase has been an intensely studied topic since the discovery of its physical implications ~\cite{Berry84,Simon83,Bohm03,Vanderbilt_book,Cohen19,KaneRMP,ZhangSCRMP,ChiuRMP}. For example, the Berry phase of pure states plays an important role in the study of topological matter since it lays the foundation for characterizing topological properties~\cite{TKNN,Haldane,KaneRMP,ZhangSCRMP,MooreN,KaneMele,KaneMele2,BernevigPRL,MoorePRB,FuLPRL,Bernevigbook,ChiuRMP}. The formalism of geometric phase can also be generalized to quantum systems at finite temperatures  described by density matrices. There have been many approaches to characterize geometric phases for mixed states \cite{Uhlmann86,Uhlmann89,Uhlmann91,GPMQS1,GPMQS2,GPMQS3,GPMQS4,GPMQS5,GPMQS6,Andersson13,DiehlPRX17,WangSR19}, among which the Uhlmann phase \cite{Uhlmann86,Uhlmann89,Uhlmann91} and the interferometric phase \cite{GPMQS1} are frequently mentioned and widely applied.

The interferometric phase proposed by Sj$\ddot{\text{o}}$qvist et al. \cite{GPMQS1} and developed in subsequent works~\cite{PhysRevA.67.020101,PhysRevLett.90.160402,Faria_2003,PhysRevLett.93.080405,GPM06,ComUI16} is built by generalizing the optical process of the Mach-Zehnder interferometer to a unitary evolution of mixed states. The interferometric phase is related to the Berry phase in the sense that the former may be viewed as a type of thermal average of the latter. The interferometric phase has been realized and measured in experiments by using nuclear magnetic resonance \cite{PhysRevLett.91.100403,GHOSH200627}, polarized neutrons \cite{PhysRevLett.101.150404} and Mach-Zehnder interferometer \cite{PhysRevLett.94.050401}. Moreover, the formalism was extended to nonunitary evolutions \cite{PhysRevA.67.020101,PhysRevLett.90.160402,Faria_2003,PhysRevLett.93.080405,GPM06}.
Meanwhile, the Uhlmann phase follows a mathematical construction similar to that of the Berry phase by developing a formalism of the density matrix, which inherits the topological nature since it reflects the holonomy when the system traverses a loop in the parameter space. It has been applied as a topological indicator to exemplary quantum systems at finite temperatures \cite{ViyuelaPRL14,ViyuelaPRL14-2,OurPRA21,Galindo21,Zhang21}. In those cases, the Uhlmann phase jumps at a critical temperature $T_c$, indicating a change of the topological structure with temperature. More recently, experimental simulations of the Uhlmann phase have been realized by controlling a bipartite entangled state formed by a system of interest and an ancilla for environmental effects~\cite{npj18}.
The Uhlmann phase can be described by the fiber-bundles language \cite{Uhlmann89,Uhlmann91,ourPRB20}. However, the principal bundle, or the Uhlmann bundle, is a trivial one \cite{TDMPRB15}. Therefore, all the characteristics, including the Chern number, vanish. Nevertheless, the Uhlmann phase corresponds to the Uhlmann holonomy and is still able to reflect the underlying topological properties.

While both interferometric and Uhlmann phases can be formulated as the phases from the evolution following the corresponding parallel transport conditions, the two geometric phases are inequivalent. The Uhlmann phase requires manipulations of the ancilla while the interferometric phase in its original form does not. Moreover, the interferometric phase does not require the concept of holonomy used in Uhlmann's formalism. In a previous comparative study of the two geometric phases of the Kitaev chain \cite{ComUI16}, it was found that the interferometric phase is intact when temperature varies, leading to the claim that there is no discrete jump of the interferometric phase at finite temperatures and accordingly, no finite-temperature topological transition.
It can be shown that the absence of any finite-temperature transition of the interferometric phase holds for all two-level systems and some other cases. Nevertheless, the interferometric phase can capture topological features of the band structure of the Kitaev chain just like the Berry phase.
In contrast, the Uhlmann phase of two-level systems already exhibits finite-temperature transitions~\cite{ViyuelaPRL14,OurPRA21,Galindo21}, including the Kitaev chain. In our later discussion, we will present an explicit example showing a finite-temperature transition of the interferometric phase in a three-level system.

To facilitate a fair comparison, we first prove that it is possible for a physical process to satisfy both parallel-transport conditions of the interferometric and Uhlmann phases, at least when unitary evolution is considered. After specifying the requirements for such processes, we also explain the resulting phase and its implications.
By examining specific two-level and three-level models following the process compatible with both parallel-transport conditions, we contrast the difference between the two geometric phases. The interferometric phase only exhibits discrete jumps at finite temperatures in the three-level model when traversing a certain type of loops in the parameter space while such finite-temperature transitions of the Uhlmann phase appear in both two-level and three-level systems. The comparison thus demonstrates the rich physics associated with topology of finite-temperature quantum systems.

The rest of the paper is organized as follows. Sec.~\ref{sec:overview} briefly reviews and compares the interferometric and Uhlmann phases via their geometric frameworks and then characterizes the class of processes that satisfy both parallel-transport conditions. Sec.~\ref{sec:example} presents exactly solvable models of two-level and three-level systems to contrast the two phases and a discussion on their physical implications. Sec.~\ref{sec:conclusion} concludes our work. The Appendix summarizes some details and derivations.

\section{Overview of two geometric phases of mixed states}\label{sec:overview}
\subsection{Purification of density matrix}
Since the frameworks of both the interferometric and Uhlmann phases can be described via purification of density matrices, we begin with a brief overview of purification here.
For simplicity, we set $\hbar=k_B=1$ in the following. In quantum mechanics, a mixed quantum state is represented by a density matrix $\rho$, a Hermitian operator carrying no phase information. To define a geometric phase similar to the Berry phase, Uhlmann introduced the concept of purification or amplitude of $\rho$ \cite{Uhlmann86,Bengtsson_book} via $W=\sqrt{\rho}V$, where $\rho$ should have full rank. $W$ and the unitary matrix $V$ play the roles of wavefunction and phase factor, respectively. Equivalently, $W$ is said to purify $\rho$ since $\rho=WW^\dag$.
Let $N$ be the rank of $\rho$.
If $\rho$ is diagonalized as $\rho=\sum_{n=0}^{N-1}\lambda_n|n\rangle\langle n|$, then $W=\sum_{n=0}^{N-1}\sqrt{\lambda_n}|n\rangle\langle n|V$. The purification $W$ is a $N\times N$ matrix that is isomorphic to a $N^2$-dimensional state-vector 
called the purified state of $\rho$. Explicitly,
\begin{align}\label{W1}
|W\rangle=\sum_n\sqrt{\lambda_n}|n\rangle_s\otimes V^T|n\rangle_a,\end{align} where $|n\rangle_s$ and $|n\rangle_a$ are respectively called system and ancilla states. The ancilla is an auxiliary system encoding the environmental effects on the system, which is defined up to a unitary transformation $V$. The introduction of the ancilla and purified state allows us to rewrite quantum statistical expressions in terms of quantum-mechanical like expressions. While suitable manipulations of the ancilla are often present in a nontrivial Uhlmann process, manipulations of the ancilla are not necessary for the interferometric phase. This subtlety will be clarified in our later discussions. 
It can be shown that the inner product between two purified states follows the Hilbert-Schmidt product
\begin{equation}\label{eq:HSP}
\langle W_1|W_2\rangle=\text{Tr}(W^\dagger_1 W_2).
\end{equation}
Moreover, the density matrix of the system can be obtained by
\begin{align}\label{rho1}
\rho=\text{Tr}_a(|W\rangle\langle W|),
\end{align}
where $\text{Tr}_a$ means the partial trace is taken over the ancilla space.
The physical meaning of $W$ is still under debate since it is a matrix but plays the similar role as the wavefunction. To simulate mixed states on classical or quantum computers, one instead constructs the purified state $|W\rangle$ by employing an ancilla state entangled with the system state \cite{Bengtsson_book,npj18}.

\subsection{Interferometric phase}\label{IIB}
The original introduction of the interferometric phase of mixed states is quite straightforward \cite{GPMQS1} without decomposing the density matrix to obtain a matrix-valued phase factor.
Inspired by the optical process of the Mach-Zehnder interferometer, Sj$\ddot{\text{o}}$qvist et al.~\cite{GPMQS1} directly assigned a phase to a mixed state after a unitary evolution $U(t)$. Explicitly, if the density matrix evolves according to $\rho(t)=U(t)\rho(0)U^\dagger(t)$, the system with density matrix $\rho(t)$ obtains a phase \begin{align}\label{thetaU}\theta=\arg\text{Tr}[\rho(0)U(t)]\end{align}
with respect to the initial state $\rho(0)$. 
If $\rho$ initially describes a pure state of the form $\rho(0)=|\psi(0)\rangle\langle\psi(0)|$, under a unitary evolution such that $|\psi(t)\rangle=U(t)|\psi(0)\rangle$, the phase between $\rho(t)=|\psi(t)\rangle\langle\psi(t)|$ and $\rho(0)$ naturally reduces to the known result $\theta=\arg\text{Tr}[\rho(0)U(t)]=\arg\langle\psi(0)|\psi(t)\rangle$ according to Eq.~(\ref{thetaU}).

The phase discussed above is general. Ref.~\cite{GPMQS1} further introduced the interferometric phase if $U(t)$ is a parallel transport, which we will briefly explain here.  
Recall that when considering pure states $|\psi_{1,2}\rangle$, if the overlap $\langle \psi_1|\psi_2\rangle $ is a positive real number, the two states are said to be `in phase' or `parallel' with each other. As a generalization, Ref.~\cite{GPMQS1} called a unitary transformation $U(t)$ a parallel transport if $\rho(t+\dif t)$ is always in phase with $\rho(t)$, meaning that $\theta=0$ according to Eq.~(\ref{thetaU}).   Note the relative transformation that takes $\rho(t)$ to $\rho(t+\dif t)$ is $U(t+\dif t)U^\dagger (t)$ since
\begin{align}\label{pfi1}
\rho(t+\dif t)
&=U(t+\dif t)U^\dagger (t)\rho(t)U(t)U^\dagger(t+\dif t).
\end{align}
Then, Eq.~(\ref{thetaU}) indicates that the `in phase' condition between $\rho(t)$ and $\rho(t+\dif t)$ is
\begin{align}\label{pfi2}
&\arg\text{Tr}\left[\rho(t)U(t+\dif t)U^\dagger (t)\right]=0,
\end{align}
which is the parallel-transport condition suggested by Sj$\ddot{\text{o}}$qvist et al.~\cite{GPMQS1}.
An expansion of the left-hand-side gives $\text{Tr}[\rho(t)U(t+\dif t)U^\dagger (t)]\approx 1+\dif t \text{Tr}[\rho(t)\dot{U}(t)U^\dagger (t)]$.
Since $\text{Tr}[\rho(t)\dot{U}(t)U^\dagger (t)]$ is an imaginary number, the condition (\ref{pfi2}) indicates
\begin{align}\label{pfi4}
 \text{Tr}\left[\rho(t)\dot{U}(t)U^\dagger (t)\right]=0.
 \end{align}
 For a pure state with $\rho(t)=|\psi(t)\rangle\langle\psi(t)|$, Eq.~(\ref{pfi4}) reduces to \begin{align}\label{pfi4b}\langle\psi(t)|\frac{\dif}{\dif t}|\psi(t)\rangle=0,\end{align} which is the parallel-transport condition for pure states \cite{AAPRL87}.

If $U(t)$ describes a dynamical evolution governed by the Hamiltonian $H$, then $\mi \dot{U}=HU$, and the condition (\ref{pfi4}) becomes
\begin{align}\label{dp0}
 \text{Tr}\left[\rho(t)H(t)\right]=0.
\end{align}
The accumulated dynamical phase during $U(t)$ is
\begin{align}\label{dp1}
\theta_D=-\int_0^t \text{Tr}\left[\rho(t')H(t')\right]\dif t'.
\end{align}
Thus, the parallel-transport condition requires that the dynamical phase vanishes.
The total phase is the sum of the dynamical and geometric phases. If the dynamical phase vanishes according to condition (\ref{pfi4}), only the geometric phase is accumulated, leading to the interferometric phase
\begin{align}\label{gp1}
\theta_I=\arg\text{Tr}\left[\rho(0)U(t)\right]=\arg\text{Tr}\left[\rho(t)U(t)\right].
\end{align}

Theoretically, the operator $U(t)$ for constructing the interferometric phase should be a solution to Eq.~(\ref{pfi4}). However, this single equation is not sufficient to fully determine $U(t)$, which is a $N\times N$ matrix. If we diagonalize $\rho(t)$ as $\rho(t)=\sum_n\lambda(t)|n(t)\rangle\langle n(t)|$, it follows from Eq.~(\ref{gp1}) that the interferometric phase is $\theta_I=\sum_n\lambda_n(t)\langle n(t)|U(t)|n(t)\rangle$, i.e., only the $N$ diagonal elements of $U(t)$ are relevant to $\theta_I$. Thus, it was suggested \cite{GPMQS1} to strengthen the parallel-transport condition (\ref{pfi4}) by \begin{align}\label{pfi6}
\langle n(t)|\dot{U}(t)U^\dagger(t)|n(t)\rangle=0,\quad n=0,1,\cdots, N-1.
\end{align}

Alternatively, the interferometric phase can be reformulated in terms of purification of the density matrices in the form $\rho(0)=W(0)W^\dag(0)$ and $\rho(t)=W(t)W^\dag(t)$. The transformation $\rho(t)=U(t)\rho(0)U^\dagger(t)$ implies $W(t)=U(t)W(0)$. In terms of purified states, this  corresponds to $|W(t)\rangle =U(t)\otimes 1|W(0)\rangle$, implying that only the evolution of the system is relevant.
By using the Hilbert-Schmidt product~\eqref{eq:HSP}, the interferometric phase is given by
\begin{align}\label{gp3}
\theta_I&=\arg\text{Tr}\left[\rho(0)U(t)\right]
=\arg\langle W(0)|W(t)\rangle,
\end{align}
which is the relative phase between the initial and final (instantaneous) states.
Similarly, the parallel-transport condition~\eqref{pfi4} can be written as
\begin{align}\label{pc3}
0&=\text{Tr}\left[\rho(t)\dot{U}(t)U^\dagger (t)\right]
=\text{Tr}\left[W^\dag(t)\dot{W}(t)\right]\notag\\
&=\langle W(t)|\frac{\dif}{\dif t}|W(t)\rangle,
\end{align}
which is also equivalent to
\begin{align}\label{pc4}
\text{Im}\langle W(t)|\frac{\dif}{\dif t}|W(t)\rangle=0
\end{align}
due to $\langle W(t)|W(t)\rangle=1$.

The unitary evolution $U(t)$ in the derivation of the interferometric phase has a clear physical meaning because during the process, the purified state evolves according to
\begin{align}\label{W2}
|W(t)\rangle=\sum_n\sqrt{\lambda_n}U(t)|n\rangle_s\otimes |n\rangle_a.
\end{align}
When compared to the Uhlmann phase that will be discussed later,
the interferometric phase may be more accessible because the Uhlmann parallel-transport condition is matrix-valued, making it challenging to interpret the meaning. Moreover, 
Eq.~(\ref{W2}) indicates that $U$ only acts on the first Hilbert space of the purified state in the interferometric phase. 

In principle, a generic expression of the interferometric phase can be  obtained for arbitrary models. Consider a quantum system initially in a mixed state $\rho(0)=\sum_n\lambda_n|n\rangle\langle n|$, and each individual pure state in the ensemble evolves under the transformation along a loop $\gamma(t)$, then
\begin{align}\label{SjU}
U(t)=\sum_{n=0}^{N-1}\me^{-\int_{0}^t\dif t'\langle n(t')|\frac{\dif}{\dif t'}|n(t')\rangle}|n(t)\rangle\langle n(t)|.
\end{align}
It can be shown that the parallel-transport condition is indeed satisfied: $\langle n(t)|\dot{U}(t)U^\dagger(t)|n(t)\rangle=0$, $n=0,1,\cdots, N-1$. At the end of the evolution, the system acquires an interferometric phase
\begin{align}\label{pfi9}
\theta_I(\gamma)=\arg\left(\sum_n\lambda_n\me^{\mi\beta_n(\gamma)}\right),
\end{align}
where
\begin{equation}\beta_n(\gamma)=\mi\oint\dif t\langle n(t)|\frac{\dif}{\dif t}|n(t)\rangle\end{equation}
is the geometric phase of the $n$th individual pure state when evolving along $\gamma(t)$.

\subsection{Uhlmann phase}
If a quantum system depends on a set of parameters $\mathbf{R}=(R_1,R_2,\cdots, R_k)$ spanning a parameter space $M$, the Hamiltonian and density matrix can be controlled externally via these parameters.
Starting with a curve $\gamma(t)\in M$, we have $\rho(t)\equiv \rho(\mathbf{R}(t))$ and its purification $W(t)\equiv W(\mathbf{R}(t))$:
\begin{align}
t\mapsto \rho(t),\quad t\mapsto W(t),\quad \rho(t)=W(t)W^\dag(t).
\end{align}
The purification is said to be parallel-transported along $\gamma$ if the length of $\gamma$, given by
\begin{align}\label{Eq:L}
L(\gamma)=\int_\gamma\sqrt{\langle\dot{W}|\dot{W}\rangle}\dif t,
\end{align}
is minimized \cite{Uhlmann95}. As explained in Appendix \ref{appa}, the parallel-transport condition is given by
\begin{align}\label{pc}
\dot{ W}W^\dag=W\dot{W}^\dag.
\end{align}
In the language of fiber bundles, $\rho(t)$ defines a closed curve in the base space, and $W(t)$ is a lift of this loop in the total space. If $W(t)$ is parallel-transported, it is said~\cite{Uhlmann86} to be a horizontal lift of $\rho(t)$.
Clearly, the parallel-transport condition (\ref{pc}) is different from Eq.~(\ref{pfi4}) in the formalism of the interferometric phase since the former is a matrix-valued equation.
Integrating both sides, we obtain the parallel condition between two amplitudes $W_{1,2}$:
\begin{align}\label{pc2}
W_1W^\dag_2=W_2W^\dag_1>0.
\end{align}
Note the parallelity relation between amplitudes is not an equivalence relation: It lacks transitivity. We also call a process equipped with the condition (\ref{pc2}) as the Uhlmann process, which has been shown to be incompatible with dynamic processes governed by the Hamiltonian~\cite{ourPRB20}.
Thus, no dynamical phase is generated in an Uhlmann process.
Meanwhile,
the parallel-transport condition of the interferometric phase also excludes the generation of any dynamical phase. 

Here we give a brief description of the Uhlmann phase via the fiber-bundle language. More details can be found in Uhlmann's original works~\cite{Uhlmann86,Uhlmann89,Uhlmann91} or more recent works \cite{2DMat15,ourPRB20,OurUB}.
Consider a cyclic process during which the amplitude of a density matrix is parallel-transported.
Let this cyclic process be parameterized by $t\in [0,\tau]$, $0\le t\le \tau$. Here `cyclic' means the initial and final density matrices are the same: $\rho(0)=\rho(\tau)$ provided $\gamma(t)$ is a closed curve. Although $W(t)$ is parallel-transported at every $t$, the initial and final amplitudes, $W(0)$ and $W(\tau)$, may not be parallel since the parallel condition (\ref{pc}) is not transitive. Explicitly, given $W(0)=\sqrt{\rho(0)}V(0)$ and $W(\tau)=\sqrt{\rho(\tau)}V(\tau)=\sqrt{\rho(0)}V(\tau)$, the violation of Eq.~(\ref{pc2}) or $W (0)W^\dag(\tau)\neq W(\tau)W^\dag(0)$ implies \begin{align}\label{V12}V(0)V^\dag(\tau)\neq V(\tau)V^\dag(0).\end{align}
 In general, the two
phase factors may be different from each other, and they are
off by an Uhlmann holonomy depending on $\gamma$: \begin{align}\label{hnu}V(\tau)=U_{\gamma}V(0),\end{align} where $U_{\gamma}$ may not be Hermitian (due to the inequality (\ref{V12})) but always unitary. As long as the parallel-transport condition (\ref{pc}) is satisfied, it can be shown that the Uhlmann holonomy is given by
\begin{align}\label{U0}
U_{\gamma}=\mathcal{P}\me^{-\oint_\gamma A_U },
\end{align}
where $\mathcal{P}$ is the path-ordering operator and $A_U=-\dif VV^\dag$ is the Uhlmann connection. In terms of the eigenvalues and eigenvectors of $\rho$, the Uhlmann connection has the form
\begin{align}\label{AU}
A_U=-\sum_{nm}|n\rangle\frac{\langle n|[\dif\sqrt{\rho},\sqrt{\rho}]|m\rangle}{\lambda_n+\lambda_m}\langle m|.
\end{align}
To quantify the difference between the initial and final purifications, we introduce the transition amplitude between the initial and final purified states
\begin{align}
\mathcal{G}_U=\langle W(0)|W(\tau)\rangle
=\text{Tr}\left[\rho(0)\mathcal{P}\me^{-\oint_\gamma A_U }\right].
\end{align}
Its argument is the famous Uhlmann phase
\begin{align}\label{thetaUb}
\theta_U=\arg\langle W(0)|W(\tau)\rangle=\arg\text{Tr}\left[\rho(0)\mathcal{P}\me^{-\oint_\gamma A_U }\right].
\end{align}

Two features of the Uhlmann process are mentioned here. Firstly, the Uhlmann process may or may not be a unitary process. In general, the amplitude evolves as
\begin{align}
W(t)=\sum_n\sqrt{\lambda_n(t)}|n(t)\rangle\langle n(t)|V(t)
\end{align}
following the Uhlmann parallel-transport condition.
The corresponding density matrix $\rho(t)=W(t)W^\dag(t)$ may not be of the form $\rho(t)=U(t)\rho(0)U^\dag(t)$ for some unitary transformation $U$ if the eigenvalues $\lambda_n(t)$ change with $t$. Typical examples of nonunitary Uhlmann processes include two-band models \cite{ViyuelaPRL14}. Meanwhile, unitary Uhlmann processes also exist, such as the spin-$j$ model \cite{OurPRA21,Galindo21} and the three-level model to be discussed later. While nonunitary generalizations of the interferometric phase have been proposed~\cite{PhysRevLett.93.080405}, we will focus on unitary processes in the following discussion. Moreover, implications of the Uhlmann connection on the fidelity of fermionic systems have been studied in a series of works~\cite{PhysRevE.77.011129,PhysRevLett.119.015702,PhysRevB.98.245141}.

Secondly, the Uhlmann parallel-transport condition is  qualitatively different from the parallel-transport condition of the interferometric phase shown in Eq.~(\ref{pc4}).
We note that Eq.~(\ref{pc}) implies $\text{Im}(\dot{ W}W^\dag)=0$. Taking the trace of both sides, we get
\begin{align}\label{pc5}
0=\text{Im}\text{Tr}(\dot{ W}W^\dag)=\text{Im}\text{Tr}(W^\dag \dot{ W})=\text{Im}\langle W(t)|\frac{\dif}{\dif t}|W(t)\rangle,
\end{align}
which is the same as Eq.~(\ref{pc4}) suggested by Sj$\ddot{\text{o}}$qvist et al.~\cite{GPMQS1}.
Superficially, it seems that the two parallel-transport conditions are similar. However, their difference is subtle but important: Eq.~(\ref{pc5}) is a weaker necessary implication of the condition (\ref{pc}) since the latter is matrix-valued with more degrees of freedom. Moreover, the equivalence between the original `in phase' condition (\ref{pfi4}) and the reformulated identity (\ref{pc4}) or (\ref{pc5}) for the interferometric phase is valid only if there is no $t$-dependent transformation acting on the ancilla. On the other hand, the ancilla transformation is ubiquitous in Uhlmann's approach. We will comment on the subtlety later. 

\subsection{Comparison between the two geometric phases in a single process}
A comparison of Eq.~(\ref{gp3}) and Eq.~(\ref{thetaUb}) shows that both interferometric and Uhlmann phases come from the relative phase between the initial and final (instantaneous) purified states. This brings forth a sequence of intriguing problems: Is there any cyclic physical process that can simultaneously satisfy the two different parallel-transport conditions  (\ref{pfi6}) and (\ref{pc})? If such a process exists, what is the relative phase between the initial and final states at the end of the evolution? Moreover, if such a process exists, it will provide a fair comparison between the Uhlmann and interferometric phases. Here we explicitly construct such processes and specify the requirements.

For a single process to simultaneously satisfy the two parallel-transport conditions, the following conditions must be satisfied. 
Firstly, the process must be cyclic, as required by the Uhlmann holonomy. Therefore, the evolution should bring the final density matrix to be the same as the initial density matrix. However, the final purified state may be different from the initial purified state and gives rise to a geometric phase. Secondly, we will focus on the interferometric phase from a system undergoing unitary evolution. Hence, the corresponding Uhlmann process is also chosen to be unitary. Accordingly, we assume a cyclic unitary process of duration $\tau$ described by
\begin{align}\label{rhot}\rho(t)=U_s(t)\rho(0)U_s^\dag(t),\end{align}
where the subscript `$s$' emphasizes that $U_s(t)$ is a system transformation.
At the end of the cyclic process, $\rho(0)=\rho(\tau)$ implies
\begin{align}\label{rhotau}[\rho(0),U_s(\tau)]=0.\end{align}
Next, we purify the density matrix as $\rho(t)=W(t)W^\dag(t)$, or conversely $W(t)=\sqrt{\rho(t)}V(t)$. The evolution (\ref{rhot}) can be satisfied by
\begin{align}\label{Wt}
W(t)&=U_s(t)W(0)U_a(t)=U_s(t)\sqrt{\rho(0)}V(0)U_a(t) \notag \\
&=\sqrt{\rho(t)}U_s(t)V(0)U_a(t).
\end{align}
with respect to $U_s(0)=U_a(0)=1$. Here $\sqrt{\rho(t)}=U_s(t)\sqrt{\rho(0)}U_s^\dag(t)$ has been applied. When compared to $W(t)=\sqrt{\rho(t)}V(t)$, the phase factor evolves as $V(t)=U_s(t)V(0)U_a(t)$, similar to $W(t)$. We remark that $U_a(t)$ is not a U$(N)$ phase factor but an ancilla transformation with the subscript `$a$'. Its meaning becomes clear by noting that the related purified state corresponding to Eq.~(\ref{Wt}) is
\begin{align}\label{Wtb}
|W(t)\rangle=\sum_n\sqrt{\lambda_n}U_s(t)|n\rangle_s\otimes (V(0)U_a(t))^T|n\rangle_a.
\end{align}

During an Uhlmann process, $W(t)$ follows the Uhlmann parallel-transport condition. By substituting Eq.~(\ref{Wt}) into the condition (\ref{pc}), $U_{s,a}$ must satisfy
\begin{align}\label{upc}
&U^\dag_a W^\dag(0)U^\dag_s\dot{U}_sW(0)U_a+U^\dag_a W^\dag(0)W(0)\dot{U}_a\notag\\=&U^\dag_a W^\dag(0)\dot{U}^\dag_s U_sW(0)U_a+\dot{U}^\dag_a W^\dag(0)W(0)U_a.
\end{align}
The validity is actually guaranteed by Eq.~(\ref{AU}). 
A proof is outlined in Appendix \ref{appb}.


To satisfy the parallel-transport condition of the interferometric phase, we emphasize that an extra $t$-dependent evolution operator $U_a(t)$ appears in the ancilla. Such a transformation on the ancilla is necessary in the Uhlmann process. 
However, the $t$-dependent transformation on the ancilla invalidates the equivalence between Eqs.~(\ref{pfi4}) and (\ref{pc5}) for the interferometric phase. 
In the original approach of Ref.~\cite{GPMQS1}, it is the system density matrix $\rho(t)$ that must be kept in phase during parallel transport, so $U_a$ is irrelevant to the evolution of $\rho(t)$ according to Eq.~(\ref{rhot}). Thus, we are allowed to follow the strengthened parallel-transport condition (\ref{pfi6}) from Ref.~\cite{GPMQS1} by replacing $U$ by $U_s$
and $|n(t)\rangle=U_s(t)|n\rangle_s$.
Moreover, by substituting Eq.~(\ref{Wtb}) into Eq.~(\ref{pc5}), we obtain
\begin{align}\label{pc7}
\text{Tr}_s\left[\rho(0)\dot{U}_sU^\dag_s\right]+\text{Tr}_a\left[\rho^T_a(0)\dot{U}_aU^\dag_a\right]=0.
\end{align}
Here $\rho_a$ is the density matrices of the ancilla. When a general transformation on the ancilla is involved, the condition (\ref{pc5}) then defines a generalization of the interferometric phase, called the generalized Berry phase, as discussed in Ref.~ \cite{OurTB}. However, this is beyond the scope of our current discussion. 

Based on the results, as long as the Uhlmann connection
\begin{align}\label{Aut}
A_U=-\dif VV^\dag=-\dif\left[U_s V(0)U_a\right]\left[U_s V(0)U_a\right]^\dag
\end{align}
satisfies Eq.~(\ref{AU}) and $U_s$ respects Eq.~(\ref{pfi6}),
it is possible that a single physical process meets both parallel-transport conditions. 
At the end of such a process of duration $\tau$, the relative phase between the initial and final states is
  \begin{align}
&\arg\langle W(0)|W(\tau)\rangle=\arg\text{Tr}\left[W^\dag(0)W(\tau)\right]\notag\\
=&\arg\text{Tr}\left[V^\dag(0)\sqrt{\rho(0)}U_s(\tau)\sqrt{\rho(0)}V(0)U_a(\tau)\right]\notag\\
=&\arg\text{Tr}\left[\rho(0)U_s(\tau)V(0)U_a(\tau)V^\dag(0)\right],
\end{align}
where Eq.~(\ref{rhotau}) has been applied in the last step.
From $V(\tau)=U_s(\tau)V(0)U_a(\tau)$ and $V(\tau)=\mathcal{P}\me^{-\oint_\gamma A_U }V(0)$ according to Eqs.~(\ref{hnu}) and (\ref{U0}), we conclude that the relative phase between the initial and final purified states is the Uhlmann phase~\eqref{thetaUb}.
This is not surprising since Eq.~(\ref{pc4}) covers Uhlmann's parallel-transport condition (\ref{pc}) but takes a different form from the condition (\ref{pfi4}) here. 
Therefore, we follow Eq.~(\ref{pfi6}) with $U$ replaced by $U_s$ to extract the interferometric phase, which is given by 
\begin{align}\label{gp1b}
\theta_I=\arg\text{Tr}\left[\rho(0)U_s(\tau)\right].
\end{align}

To end this section, we present a comparison between the two geometric phases of mixed states. The major difference comes from the ancilla evolution $U_a$. Given an arbitrary unitary transformation $U_a$, the density matrix remains the same: $\rho(t)=W(t)W^\dag(t)=\text{Tr}_a(|W(t)\rangle\langle W(t)|)$. This means at each point $\rho(t)$ on the loop $\gamma$ of evolution, there is a corresponding linear space generated by $U_a$, which is the fiber space at $\rho(t)$. If $U_a$ and $U_s$ respect Eq.~(\ref{upc}), Uhlmann's parallel-transport condition is satisfied, and a point in the fiber space that corresponds to $W(t)$ lies on the horizontal lift of $\gamma$. Thus, the Uhlmann phase is naturally connected to the topological structure of the Uhlmann bundle via the concept of Uhlmann holonomy, and its jump signals a topological phase transition of the system. On the other hand, only the system evolution $U_s$ is relevant in the parallel-transport condition of the interferometric phase according to Sj$\ddot{\text{o}}$qvist et al.'s approach~\cite{GPMQS1}. To our knowledge, the formalism of the interferometric phase does not naturally connect to a fiber-bundle description like the Uhlmann phase. Nevertheless, the interferometric phase, as given by Eq.~(\ref{pfi9}), may be considered as the thermal average of the Berry phase factor from each energy level and thus reflects the geometrical properties of the Berry holonomy. Accordingly, we refer to the quantized jump of the interferometric phase as a geometric phase transition, which will be discussed in the following section. 

\section{Examples and discussions}\label{sec:example}
Here we analyze some concrete examples that will generate both the interferometric and Uhlmann phases and compare their behavior.

\subsection{Interferometric phase}
We begin with a simple two-level system (or a qubit) described by
\begin{equation}\label{eq:H2}
H_2=\boldsymbol{\sigma}\cdot\mathbf{R},
\end{equation}
where $\mathbf{R}=R(\sin\theta\cos\phi,\sin\theta\sin\phi,\cos\theta)^T$ and $\boldsymbol{\sigma}=(\sigma_x, \sigma_y, \sigma_z)^T$ is the vector of the Pauli matrices. At temperature $T$, the canonical-ensemble density matrix is $\rho=\frac{1}{Z}\me^{\beta H_2}=\frac{1}{2}\left[1-\tanh(\beta R)\boldsymbol{\sigma}\cdot\hat{\mathbf{R}}\right]$, where $\beta=\frac{1}{T}$, $R=|\mathbf{R}|$, and $\hat{\mathbf{R}}=\frac{\mathbf{R}}{R}$.
Applying Eq.~(\ref{pfi9}), we obtain the interferometric phase
\begin{align}\label{pfi9b}
\theta_I(\gamma)=\arctan\left[\tanh(\beta R)\tan(\beta_-(\gamma))\right],
\end{align}
where $\beta_-(\gamma)=\frac{1}{2}\oint_\gamma(1-\cos\theta)\dif \phi$ is the geometric phase of the ground state.
Here $\gamma(t)$ denotes a unitary-evolution loop on the parameter space-the two-dimensional unit sphere $S^2$. For example, suppose $\mathbf{R}$ initially points along the $x$-axis, then $H_2(t=0)\equiv H_2(\theta(0)=\frac{\pi}{2},\phi(0)=0)=\sigma_x R$. By gradually changing the direction of $\mathbf{R}$ along the equator, the Hamiltonian evolves as $H_2(\theta,\phi)=U_z(\phi)\sigma_xRU^\dag_z(\phi)=R(\sigma_x\cos\phi+\sigma_y\sin\phi)$, where $U_z(\phi)=\me^{-\frac{\mi}{2}\phi\sigma_z}$ is a $\phi$-rotation about the $z$-axis.

Accordingly, the mixed state evolves unitarily as $\rho=\frac{1}{Z}\me^{-\beta U_z(\phi)H_2(0)U^\dag_z(\phi)}=\frac{1}{Z}U_z(\phi)\me^{-\beta H_2(0)}U^\dag_z(\phi)$. Here the partition function $Z$ is invariant under a unitary transformation. At the end of the evolution ($t=\tau$), $\phi=2\pi$, and $\rho(\tau)$ returns to $\rho(0)$, corresponding to a closed loop for the density matrix. The interferometric phase $\theta_I$ that the mixed state obtains with respect to the evolution can be inferred from Eq.~\eqref{gp1}. 
The expression (\ref{pfi9b}) suggests that $\theta_I$ for the two-level system is continuous with respect to $\beta$ except at $\beta=0$ (or $T=\infty$). In other words, the interferometric phase of the two-level system has no discrete jumps provided the system does not reach the maximally
mixed state at infinite temperature. We remark that the absence of finite-temperature transition of the interferometric phase in two-level systems is a general feature~\cite{ComUI16}. On the contrary, it will be shown later the Uhlmann phase of the same two-level system already exhibits finite-temperature transitions.

The lack of finite-temperature transitions of the interferometric phase in the literature raises an interesting question: Is it possible to have temperature-induced quantized jumps of the interferometric phase at finite temperatures? Here we provide an affirmative answer by modifying the two-level model to a three-level system with the Hamiltonian
\begin{align}\label{H3l}
H&=\left (
\begin{array}{cc}
\mathbf{R}\cdot\boldsymbol{\sigma} & \\
  & R\\
\end{array}
\right )=R\left (
\begin{array}{ccc}
\cos\theta & \sin\theta\me^{-\mi\phi} & \\
 \sin\theta\me^{\mi\phi} & -\cos\theta &\\
   &  &  1
\end{array}
\right ).
\end{align}
where again $\mathbf{R}=R(\sin\theta\cos\phi,\sin\theta\sin\phi,\cos\theta)^T$. We use the convention that vanishing elements of a matrix will not be explicitly shown. The Hamiltonian can be diagonalized as $H=R\mathcal{U}(\theta,\phi) A   \mathcal{U}^{\dagger}(\theta,\phi)$, where
\begin{align}\label{U3l}
\mathcal{U}(\theta,\phi)=\left (
\begin{array}{ccc}
\cos\frac{\theta}{2} & \sin\frac{\theta}{2} & 0 \\
 \sin\frac{\theta}{2}\me^{\mi\phi} & -\cos\frac{\theta}{2}\me^{\mi\phi} & 0 \\
  0 & 0 &  1
\end{array}
\right ),
A=\left (
\begin{array}{ccc}
1 &  & \\
  & -1 &\\
   &  &  1
\end{array}
\right ) .
\end{align}
The Hamiltonian has two eigenvalues $\pm R$ with $+R$ being doubly degenerate. The associated eigenstates are
\begin{align}\label{EL}
|+R_{1}\rangle&=\left (
\begin{array}{c}
\cos\frac{\theta}{2}  \\
 \sin\frac{\theta}{2}\me^{\mi\phi} \\
0
\end{array}
\right ), \quad
|+R_{2}\rangle=\left (
\begin{array}{c}
0  \\
0 \\
1
\end{array}
\right ),\notag\\
|-R\rangle&=\left (
\begin{array}{c}
\sin\frac{\theta}{2}  \\
 -\cos\frac{\theta}{2}\me^{\mi\phi} \\
0
\end{array}
\right ).
\end{align}
The parameter space is also $S^2$, whose local coordinates are $(\theta,\phi)$. A loop in $S^2$ can be parameterized by $(\theta(t),\phi(t))$ with $0\le t\le\tau$ such that $\theta(0)=\theta(\tau)$, $\phi(0)=\phi(\tau)$. Hence, $\mathcal{U}$ itself induces a unitary transformation $\mathcal{U}(\theta(t),\phi(t))$. Note
\begin{align}
\mathcal{U}^\dag\dot{\mathcal{U}}&=\left (
\begin{array}{ccc}
0 & \frac{1}{2} & 0 \\
-\frac{1}{2} & 0 & 0 \\
  0 & 0 &  0
\end{array}
\right )\dot{\theta}\notag\\&+
\left (
\begin{array}{ccc}
\mi\sin^2\frac{\theta}{2} & -\mi\sin\frac{\theta}{2}\cos\frac{\theta}{2} & 0 \\
-\mi\sin\frac{\theta}{2}\cos\frac{\theta}{2} & \mi\cos^2\frac{\theta}{2} & 0 \\
  0 & 0 &  0
\end{array}
\right )\dot{ \phi}.
\notag
\end{align}
A straightforward evaluation shows
\begin{align}
\text{Tr}(\rho\dot{\mathcal{U}}\mathcal{U}^\dag)
&=\frac{1}{Z}\text{Tr}\left(\mathcal{U}\me^{-\beta R A}\mathcal{U}^\dagger\dot{\mathcal{U}}\mathcal{U}^\dagger\right)\notag\\
 &=\frac{2\mi}{Z}\dot{\phi}\left[\sinh(\beta R)\sin^2\frac{\theta}{2}-\frac{1}{2}\me^{\beta R}\right].
\end{align}
Thus, the allowed parallel-transport at finite temperature must be along circles of longitude (meridians), i.e., $\dot{\phi}=0$.

However, a subtlety arises here due to the traditional choice of the ranges of the spherical coordinates given by $0\le \theta\le \pi$, $0\le \phi<2\pi$. There are two superficial singular points in this coordinate system, the north and south poles, at which the latitudes are $\theta=0$ and $\pi$, respectively, causing the longitudes not well defined. When traversing a circle of longitude with $\phi=\phi_0$, the longitude suddenly jumps to $\phi_0\pm\pi$ after passing the south or north pole.
In fact, a whole meridian contains two semi-meridians: $(0\le \theta<\pi,\phi_0)$ with longitude $\phi_0$ and $(\pi\ge \theta>0,\phi_0+\pi)$ with longitude $\phi_0+\pi$. This artifact leads to artificial singularities in $\dot{\phi}=0$ for the parallel-transport condition at the two poles. Fortunately, those spurious singularities can be avoided by making a slight adjustment to the ranges of the spherical coordinates. In Eq.~(\ref{H3l}), the Hamiltonian is invariant under the symmetry transformation $(\theta,\phi)\rightarrow(2\pi-\theta,\phi+\pi)$, which maps a semi-meridian of longitude $\phi$ to that of $\phi+\pi$. Thus, we can redefine the ranges of the spherical coordinates as $0\le \theta<2\pi$, $0\le \phi\le \pi$. In this convention, a whole circle of longitude $\phi_0$ is expressed as $(0\le \theta<2\pi,\phi_0)$, which covers without singularity the two semi-meridians in the previous definition.

\begin{figure}[t]
\centering
\includegraphics[width=3.3in, clip]{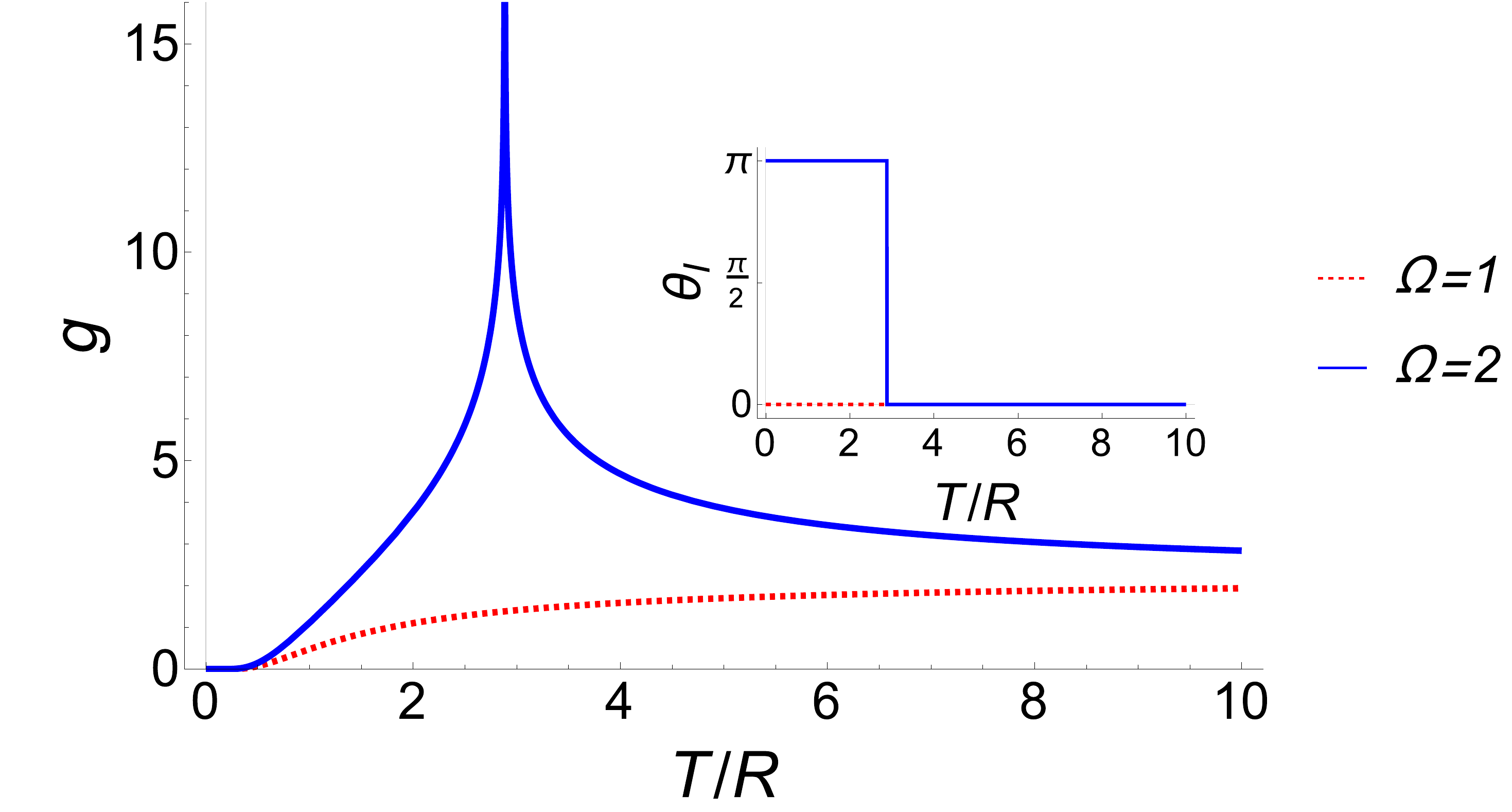}
 \caption{Geometrical generating function and interferometric phase (inset) of the model \eqref{H3l} as functions of temperature. The red dotted and blue solid lines
correspond to $\Omega=1$ and $2$, respectively. When $\Omega$ is even, $g$ diverges at $T_c$ and $\theta_I$ has a discrete jump.}
 \label{Fig1}
\end{figure}

Suppose the system initially stays at the north pole of $S^2$ and starts to evolve along a meridian of longitude $\phi_0$ at $t=0$. We also assume that the evolution path can slightly deviate from a certain meridian such that the winding number of the path may be higher than one by making multiple rounds on $S^2$.
To ensure the evolution path will not introduce a non-negligible contribution to $\dot{\phi}$, we assume the total deviation $|\dif\phi|$ along a whole meridian is at most comparable to $|\dif\theta|$ along a small segment $\dif t$. Thus, $|\dot{\phi}|$ is a higher order infinitesimal compared to $|\dot{\theta}|$. 
We define $\Omega=\frac{1}{2\pi}\oint\dif \theta$ as the number of the evolution path circling a meridian. The initial density matrix is $\rho(0)=\frac{1}{Z(0)}\text{diag}(\me^{-\beta R},\me^{\beta R},\me^{-\beta R})$. At the end of the evolution, the transformation $\mathcal{U}$ is
\begin{align}\label{Uf}
\mathcal{U}(2\pi\Omega,\phi_{0})=\left (
\begin{array}{ccc}
\cos(\pi\Omega) & 0 & \\
0 & -\cos(\pi\Omega) &\\
   &  &  1
\end{array}
\right ),
\end{align}
which is independent of the longitude $\phi_0$.
  Thus, the interferometric phase for any meridian is given by
  \begin{align}
&\theta_I(T)=\text{arg}\text{Tr}\left[\rho(0)\mathcal{U}(2\pi\Omega,\phi_{0})\right] \notag\\
=&\arg\frac{\cos(\pi\Omega)\me^{-\beta R}+\me^{-\beta R}-\cos(\pi\Omega)\me^{\beta R}}{Z(0)}.
\end{align}
Importantly, when $\Omega$ is even, one can see that $\theta_I(T)$ has a quantized jump at the critical temperature $T_c=\frac{2}{\ln2}R$ independent of $\Omega$.
Explicitly,
  \begin{align}\label{Tc3}
\theta_I(T)=\left\{\begin{array}{cc} \pi, & T<T_c,\\ 0, & T>T_c.\end{array}\right.
\end{align}

To characterize the features at $T_c$, we introduce the geometrical generating function \cite{OurPRB20b}
\begin{align}g=-\lim_{L\rightarrow\infty}\frac{1}{L}\ln|\mathcal{G}_I(T)|^2,\end{align}  where $L$ is the degrees of freedom of the system, and $\mathcal{G}_I(T)=\langle W(0)|W(\tau)\rangle$ with its norm called the visibility by analogy with optical process \cite{GPMQS1}. A jump of $\theta_I$ indicates that $T_c$ is a zero of the visibility, which indicates that the initial and final purified states are orthogonal even though the initial and final density matrices are the same in a cyclic process. 
Furthermore, $g$ has non-analytical behavior at $T_c$. This is quite similar to the dynamical quantum phase transition of quantum quench processes \cite{DQPTreview18} because $g$ is the counterpart of the dynamical free-energy density, and $|\mathcal{G}_I(T)|$ is the counterpart of the Loschmidt echo.

We visualize our findings in Fig. \ref{Fig1}, where $g$ and $\theta_I$ are plotted as functions of $T$ for $\Omega=1$ and $2$, respectively. When $\Omega=1$ (odd), $g$ varies continuously as $T$ increases, and $\theta_I$ is trivial at any finite $T$. When $\Omega=2$ (even), $\theta_I$ jumps from $\pi$ to $0$ as the temperature increases across $T_c$. Moreover, $g$ exhibits non-analytical behavior at $T_c$, signaling a change of the geometrical nature of the system. By comparing with the dynamical quantum phase transition after a quantum quench~\cite{DQPTreview18}, the behavior of the $\Omega=2$ case may be recognized as a phase transition since temperature can be thought of as the complex continuation of time and the geometrical generating function plays the role of free energy. We call the transition shown in Fig.~\ref{Fig1} a geometric phase transition since it is signaled by a jump of the interferometric phase, which is a geometric phase of mixed states.

We elaborate on the physical meaning of $T_c$ from a jump of $\theta_I$. For a system in thermal equilibrium, Eq.~(\ref{pfi9}) indicates that $\theta_I$ comes from a thermal average of the Berry phase factors.
In the zero-temperature limit, $\lim_{T\rightarrow 0}\frac{\lambda_{n>0}}{\lambda_0}=\lim_{\beta\rightarrow\infty}\me^{-\beta(E_{n>0}-E_0)}=0$, so the contribution is solely from the ground state. Thus, $\theta_I$ reduces to the corresponding Berry phase as $T\rightarrow 0$, which is consistent with the argument that the interferometric phase inherits the geometrical properties of the Berry phase at low temperatures~\cite{KitaevChain2016Sj}. In the infinite-temperature limit, $\rho$ corresponds to the maximally mixed state with the thermal weight of each level being equal. Accordingly, $\theta_I$ loses its resemblance to the ground-state Berry phase. Hence, the interferometric phase can be thought of as a measure to detect the temperature where $\rho$ loses its ability to capture the ground-state geometrical properties.

For the simplest two-level system studied above, one may assign $T_c=\infty$ according to Eq.~(\ref{pfi9b}). This implies that $\theta_I$ is relatively insensitive to temperature in the two-level case. Nevertheless, it has been proposed that driven by other parameters, $\theta_I$ may reflect the same phase transition  just as the Berry phase does, as illustrated in a study on the Kitaev chain~\cite{KitaevChain2016Sj}.
In contrast, the three-level system studied above indeed shows that $T_c$ can be finite. Therefore, $\theta_I$ resembles the Berry phase of the ground state when $T<T_c$ but changes to resemble the Berry phase of the excited states when $T>T_c$. This is corroborated by Eq.~(\ref{Uf}), where $-\cos(\pi\Omega)$ and $\cos(\pi\Omega)$ are from the Berry phase factors of the ground state and the parameter-dependent excited state, respectively. 
One may envision that there might exist more than one transition points at finite temperatures for more complicated multiple-level systems according to the interferometric phase.  

\begin{figure}[t]
\centering
\includegraphics[width=3.3in, clip]{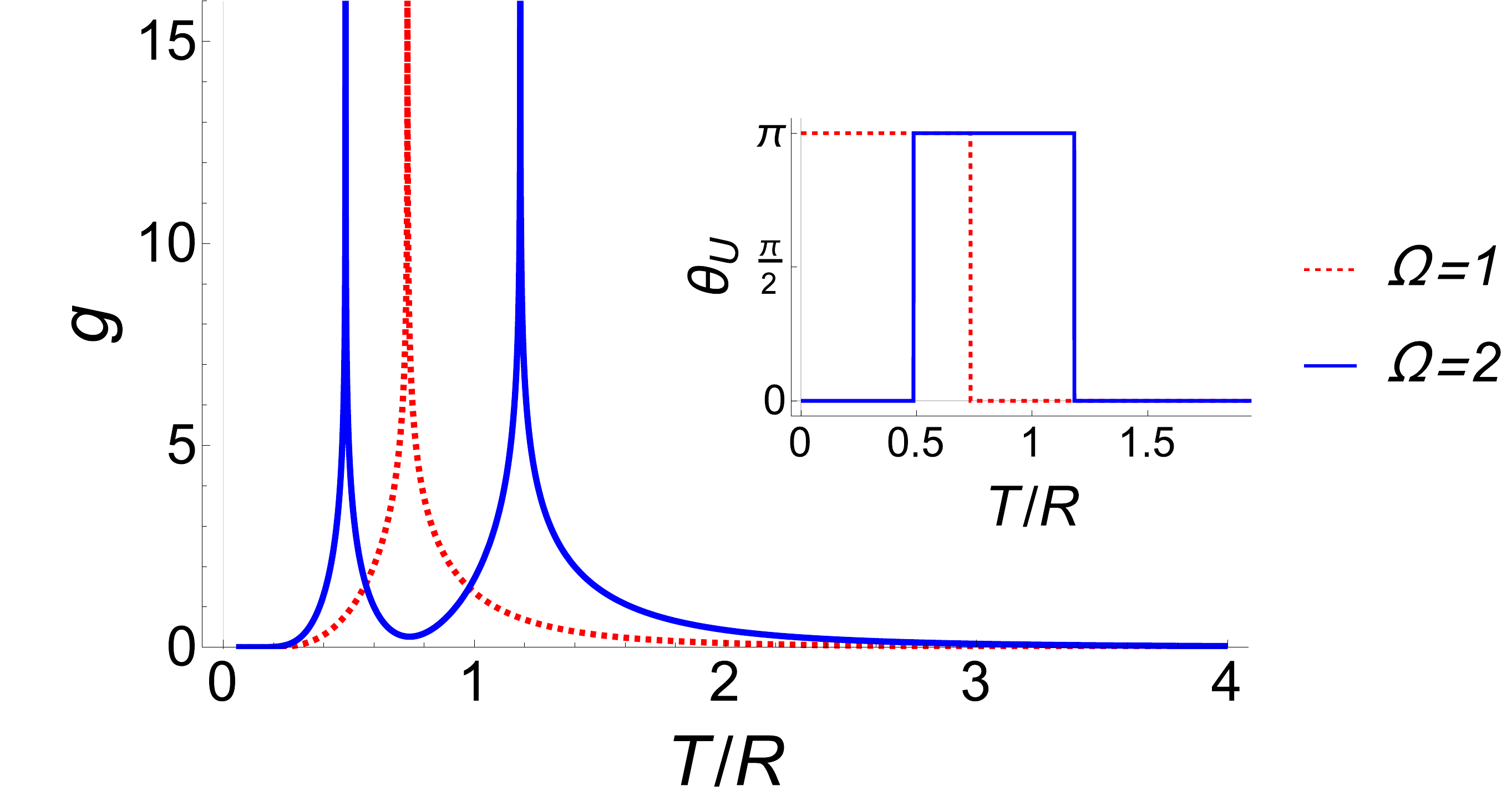}
 \caption{Geometrical generating function and Uhlmann phase (inset) of the model \eqref{H3l} as functions of temperature. The red dotted and
blue solid lines correspond to $\Omega=1$ and $2$, respectively. $g$ exhibits non-analytical behavior at topological phase transitions with quantized jumps of the Uhlmann phase.}
 \label{Fig2}
\end{figure}

\subsection{Uhlmann phase}
As a comparison, we also study the Uhlmann phase using the same models with the Uhlmann parallel-transport condition satisfied. We have given a brief discussion of the Uhlmann phase of similar models in our previous work \cite{OurPRB20b} but using a different calculation. Here we will present a detailed study to contrast the result with the interferometric phase.
A similar example to the two-level system shown in Eq.~\eqref{eq:H2} is the spin-$\frac{1}{2}$ model undergoing a unitary Uhlmann process, which has been discussed in our previous work \cite{OurPRA21}. Here we quote some key results by identifying the parameters of the Hamiltonian ~\eqref{eq:H2} with $\mathbf{R}=\mu_B\mathbf{B}$, 
where $\mu_B$ is the Bohr magneton and $\mathbf{B}$ is the external magnetic field.
The parameter space of this model is also $S^2$ and the great circles on $S^2$, such as the equator and meridians, are evolution loops satisfying Uhlmann's parallel-transport condition. When evolving along one of those loops, the system obtains an Uhlmann phase $\theta_U=\arg\left[\cos(\pi\Omega)\cos\left(\pi\Omega\text{sech}\frac{\beta R}{2}\right)\right]$ where $\Omega$ is the winding number counting how many times the loop wraps around a great circle in the parameter space.
Different from the interferometric phase shown in Eq.~\eqref{pfi9b}, the Uhlmann phase of the two-level system already exhibits quantized jumps at $T_c=\frac{R}{2\ln(\frac{\Omega}{n+\frac{1}{2}}+\sqrt{(\frac{\Omega}{n+\frac{1}{2}})^2-1})}$, where $n=0,1,\cdots, \Omega-1$. For the Uhlmann phase, a jump signifies a topological change of the Uhlmann holonomy. 

Next, we consider the same three-level model described by Eq.~(\ref{H3l}).
Under the evolution $\mathcal{U}(t)$, it is straightforward to get
\begin{align}
[\dif\sqrt{\rho},\sqrt{\rho}]
=\frac{\{\dif \mathcal{U}\mathcal{U}^{\dagger},\me^{-\beta \hat{H}}\}}{Z}+\frac{2\me^{-\frac{\beta H}{2}}\mathcal{U}\dif \mathcal{U}^{\dagger}\me^{-\frac{\beta H}{2}}}{Z} .
\end{align}
Thus, the Uhlmann connection is given by
\begin{align}\label{AUsj3l}
A_U=\sum_{m,n=+1,+1,-1}\chi_{mn}|\psi_{m}\rangle\langle\psi_{m}|\mathcal{U}\dif \mathcal{U}^{\dagger}|\psi_{n}\rangle\langle\psi_{n}|,
\end{align}
where $\chi_{mn}=\frac{
\me^{-m\beta R}+\me^{-n\beta R}-2\me^{-\frac{m+n}{2}\beta R}}
{\me^{ -m\beta R}+\me^{-n\beta R}}$,
$|\psi_n\rangle=|+R_1\rangle$, $|+R_2\rangle$ or $|-R\rangle$. Using
\begin{eqnarray}\label{dU3l}
\dif \mathcal{U}^{\dagger}(\theta,\phi)&=&\left(
\begin{array}{ccc}
-\frac{1}{2}\sin\frac{\theta}{2} & \frac{1}{2}\cos\frac{\theta}{2}\me^{-\mi\phi} & 0 \\
 \frac{1}{2}\cos\frac{\theta}{2} & \frac{1}{2}\sin\frac{\theta}{2}\me^{-\mi\phi} & 0 \\
  0 & 0 &  0
\end{array}
\right)\dif\theta\notag\\
&+&
\left (
\begin{array}{ccc}
0 & -\mi\sin\frac{\theta}{2}\me^{-\mi\phi} & 0 \\
0 & \mi\cos\frac{\theta}{2}\me^{-\mi\phi} & 0 \\
  0 & 0 &  0
\end{array}
\right) \dif\phi.
\end{eqnarray}
It follows that
\begin{align}
A_U&=\frac{\chi}{2}
\left (
\begin{array}{ccc}
0 & -\me^{-\mi\phi} & 0 \\
\me^{\mi\phi} & 0 & 0\\
 0  & 0 &  0
\end{array}
\right )  \dif\theta\notag\\&-\frac{\mi \chi}{2}
\left (
\begin{array}{ccc}
-\sin\theta & \cos\theta\me^{-\mi\phi} & 0\\
\cos\theta\me^{\mi\phi}& \sin\theta & 0 \\
  0 & 0 &  0
\end{array}
\right ) \sin\theta \dif\phi,
\end{align}
where $\chi\equiv\chi_{1,-1}=\chi_{-1,1}=\frac{
\me^{-\beta R}+\me^{\beta R}-2}
{\me^{ -\beta R}+\me^{\beta R}}$ is the only nonzero component of $\chi_{mn}$.
When the system evolves along a meridian of longitude $\phi=\phi_0$, $\dot{\phi}=0$, and the corresponding Uhlmann holonomy is
\begin{align}
&\mathcal{P}\me^{-\oint A_U}=\left (
\begin{array}{ccc}
\cos(\Omega\pi \chi) & \sin(\Omega\pi \chi)\me^{-\mi\phi_0} & \\
-\sin(\Omega\pi \chi)\me^{\mi\phi_0} & \cos(\Omega\pi \chi) &\\
   &  &  1
\end{array}
\right ).
\end{align}
At the end of evolution, the system acquires an Uhlmann phase given by Eq.~\eqref{thetaUb}. Explicitly,
\begin{align}
\theta_U=\text{Tr}[\rho(0)\mathcal{P}\me^{-\oint A_U}]=\arg\mathcal{G}_U(T),\end{align}
where
\begin{align}\mathcal{G}_U(T)=\left[\frac{(-1)^\Omega2\cosh(\beta R)\cos\left( \frac{\Omega\pi
}
{\cosh(\beta R)}\right)+\me^{-\beta R}}{Z(0)}\right],
\end{align}
independent of $\phi_0$.
A comparison with the discussion of the interferometric phase shows that the norm of $\mathcal{G}_U(T)$ can also be recognized as the visibility. Accordingly, we introduce the geometrical generating function $g=-\lim_{L\rightarrow\infty}\frac{1}{L}\ln|\mathcal{G}_U(T)|^2$. A jump of $\theta_U$ also corresponds to a zero of the visibility, which indicates orthogonality between the initial and final purified states even though the initial and final density matrices are the same in a cyclic process. Moreover, $g$ diverges at a topological transition point when the Uhlmann phase jumps. 

Our numerical calculations show that $\mathcal{G}_U(T)$ always has at least one zero no matter $\Omega$ is odd or even. For example, if $\Omega=1$, there is a zero of $\mathcal{G}_U(T)$ at $T_c\approx 0.7338R$. Similarly, the value of $\theta_U$ jumps at $T_c$ where $g$ diverges, which is visualized in Fig.~\ref{Fig2}. Since $\theta_U$ reflects the topological nature of the system at finite temperatures via the Uhlmann holonomy~\cite{Uhlmann86,2DMat15}, 
$T_c$ signals a topological phase transition. Moreover, Fig.~\ref{Fig2} indicates that the winding number has a nontrivial effect on the topological phase transition in this case. If $\Omega=1$, the system is in the topologically-nontrivial phase with $\theta_U=\pi$ at low temperature $T<T_c$. Above $T_c$, the system becomes topologically-trivial with $\theta_U=0$. This is because the thermal distribution changes the topology of $W(t)$, the horizontal lift of $\rho(t)$ \cite{Uhlmann86}. If $\Omega=2$, the system experiences two distinct topological phase transitions since $\mathcal{G}_U(T)$ has two zeros. As the temperature increases from $T=0$, the system begins with the topologically-trivial phase, then jumps to the nontrivial phase, and then jumps back to the trivial phase, showing an intermediate-temperature topological regime sandwiched by trivial regimes at lower and higher temperatures~\cite{OurPRA21,Zhang21}.

We mention that the behavior of the Uhlmann phase of the three-level system is somewhat similar to the spin-$\frac{1}{2}$ system previously investigated in Ref.~\cite{OurPRA21}. The resemblance with the two-level spin-$\frac{1}{2}$ system is because the three-level model here is obtained by including a parameter-independent energy-level as indicated by Eq.~(\ref{EL}). Nevertheless, the three-level model is not a trivial generalization of a two-level system since the interferometric phase clearly shows an intrinsic difference between the two-level and three-level systems.

\subsection{Implications}
While the interferometric phase has been measured via different experimental techniques~\cite{PhysRevLett.91.100403,PhysRevLett.94.050401,PhysRevLett.101.150404,GHOSH200627}, the Uhlmann phase of a two-level system has been simulated and measured on quantum computers~\cite{npj18}. The reason that the interferometric phase can be measured from the evolution of a natural system while the Uhlmann phase is generated from an entangled state of the system and ancilla is because the transformations of the former is on the system only but there are both system and ancilla transformations for the latter, as explained in our previous discussions. We remark that previous experimental measurements of the two phases are all on two-level systems. Therefore, our analysis of the three-level system offers testable predictions, such as the discrete jump of the interferometric phase of a three-level system at finite temperatures that is absent in two-level systems, for future experiments.

Moreover, a three-level system may be represented by a system with spin 1, and the interferometric phase may be measured using the same procedure as that of a two-level system. In contrast, a three-level system on a quantum computer may be represented by two qubits \cite{OurPRA21} or using a three-state qutrit. The corresponding purified states need to be constructed for the measurement of the Uhlmann phase. Therefore, despite the possibility of satisfying both parallel-transport conditions in a single process, one may still need to construct different experiments for the two phases due to their specific requirements of physical systems. For example, in the measurement of the interferometric phase via nuclear magnetic resonance~\cite{PhysRevLett.91.100403}, no extra manipulations were needed for the ancilla since the transformation on the system alone is sufficient. In contrast, time evolution governed by engineered Hamiltonians of both the system and ancilla was implemented in the simulation of the Uhlmann phase on quantum computers~\cite{npj18}. Nevertheless, our analysis shows the conditions for a process to satisfy both parallel-transport conditions, which will allow future experiments to facilitate fair comparisons of the two geometric phases of mixed states.

\section{Conclusion}\label{sec:conclusion}
The inequivalent parallel-transport conditions of the interferometric and Uhlmann phases clearly show that while the two geometric phases of mixed states generalize the Berry phase of pure states, they have different physical requirements and implications. The class of physical processes satisfying both parallel-transport conditions analyzed here not only offers a fair comparison of the two phases and their phase transitions but also provides deeper insights into the meanings of geometric phases of mixed states. Furthermore, realizations and measurements of the two-level and three-level systems for both phases in quantum simulators or computers will help us navigate the complex web of geometry, topology, quantum physics, and temperature.

\begin{acknowledgments}
H. G. was supported by the National Natural Science Foundation
of China (Grant No. 12074064). C. C. C. was supported by the National Science Foundation under Grant No. PHY-2011360. We thank Prof. D. M. Tong for valuable discussions. 
\end{acknowledgments}

\appendix
\section{Details of Uhlmann parallel-transport condition }\label{appa}
We begin with Eq.~\eqref{Eq:L} and rewrite it as $\gamma(t)$ can be modified as
\begin{align}\label{Lc}
L(\gamma)=\int_\gamma\sqrt{\langle\dot{W}|\dot{W}\rangle}\dif t=\int_\gamma\sqrt{\text{Tr}(\dot{W}^\dag\dif t \dot{W}\dif t )}.
\end{align}
To minimize $L(\gamma)$, it is equivalent to search all possible $W(t)$ to minimize
\begin{align}\label{L2}
&\text{Tr}(\dot{W}^\dag\dif t \dot{W}\dif t )
\notag\\
\approx&\text{Tr}\left[(W(t+\dif t)-W(t))(W^\dag(t+\dif t)-W^\dag(t))\right]\notag\\
=&2-\text{Tr}\left[W(t+\dif t)W^\dag(t)+W(t)W^\dag(t+\dif t)\right].
\end{align}
Let $W_1=W(t)$ and $W_2=W(t+\dif t)$. We are set to find the maximum of $\text{Re}(W_2W^\dag_1)$. By definition, $W_1$ and $W_2$ are both of full rank, so $A\equiv W_2W^\dag_1$ is also of full rank. Thus, it has a unique polar decomposition $A=|A|V_A$, where $|A|=\sqrt{AA^\dag}$. The following inequality is valid:
\begin{align}\label{cLc}
\text{Re}\left(\text{Tr}A\right)&\le|\text{Tr}A| =|\text{Tr}(\sqrt{|A|}\sqrt{|A|}V_A)| \notag\\
&\le\sqrt{\text{Tr}|A|\text{Tr}(V^\dag_A|A|V_A)}=\text{Tr}|A|,
\end{align}
where the Cauchy-Schwartz inequality $\text{Tr}(A^\dag B)\le\sqrt{\text{Tr}(A^\dag A)\text{Tr}(B^\dag B)}$ has been applied.
The inequality (\ref{cLc}) is saturated if $\sqrt{|A|}=\sqrt{|A|}V_A$, i.e., $V_A=1$, which implies \begin{align}\label{A}
A=\sqrt{AA^\dag}>0,
\end{align}
where `$>0$' means all eigenvalues of the corresponding
matrix are positive. This is because the full-ranked matrix $\sqrt{AA^\dag}$ only has positive eigenvalues.
Eq.~(\ref{A}) further leads to $A^2=AA^\dag$ or $A=A^\dag$, which is
\begin{align}\label{PC2}
 W(t+\dif t)W^\dag(t)=W(t)W^\dag(t+\dif t)>0.
\end{align}
Expanding both sides to the first order, we get the parallel-transport condition
\begin{align}
\dot{ W}(t)W^\dag(t)=W(t)\dot{W}^\dag(t).
\end{align}

\section{Proof of Eq.~(\ref{upc})}\label{appb}

For simplicity, we assume the initial phase factor is trivial: $V(0)=1$. Thus, $W(0)=\sqrt{\rho(0)}$, $U(t)=U_s(t)U_a(t)$, and Eq.~(\ref{upc}) becomes
\begin{align}
&U^\dag_a \sqrt{\rho_0}U^\dag_s\dot{U}_s\sqrt{\rho_0}U_a+U^\dag_a \rho_0\dot{U}_a\notag\\=&U^\dag_a \sqrt{\rho_0}\dot{U}^\dag_s U_s\sqrt{\rho_0}U_a+\dot{U}^\dag_a \rho_0U_a,
\end{align}
where $\rho_0\equiv \rho(0)$. 
This equality can be further rearranged as
\begin{align}\label{upc2}
&U_s\sqrt{\rho_0}U^\dag_s\dot{U}_s\sqrt{\rho_0}U^\dag_s-U_s\sqrt{\rho_0}\dot{U}^\dag_sU_s\sqrt{\rho_0}U^\dag_s\notag\\
=&U_sU_a\dot{U}_a^\dag\rho_0U^\dag_s-U_s\rho_0 \dot{U}_a U_a^\dag U^\dag_s.
\end{align}

Now we show that the modified parallel-transport condition (\ref{upc2}) in a unitary Uhlmann process can be inferred from Eq.~(\ref{AU}). Since $V(0)=1$, $A_U=-\dif (U_sU_a) (U_sU_a)^\dag$. Suppose $X$ is the tangent vector of the closed curve $\rho(t)$, then
\begin{align}\label{AUb}
A_U(X)=-\dot{U}_sU^\dag_s-U_s\dot{U}_aU^\dag_a U^\dag_s.
\end{align}
Moreover, Eq.~(\ref{AU}) is equivalent to
\begin{align}\label{AU2}
\rho A_U+A_U\rho=[\sqrt{\rho},\dif \sqrt{\rho}].
\end{align}
Let $\tilde{X}$ be the horizontal lift of $X$, then $\dif \sqrt{\rho}(\tilde{X})=\dot{\sqrt{\rho}}$. Therefore, Eq.~(\ref{AU2}) leads to ~\cite{ourPRB20}
\begin{align}\label{AU3}
\rho A_U(X)+A_U(X)\rho=[\sqrt{\rho},\dif \sqrt{\rho}(\tilde{X})]=[\sqrt{\rho},\dot{\sqrt{\rho}}].
\end{align}
Substituting $\sqrt{\rho}=U_s\sqrt{\rho_0}U_s^\dag$ into the right-hand-side of Eq.~(\ref{AU3}) and rearranging terms, we get
\begin{align}\label{upc3}
&U_s\sqrt{\rho_0}U^\dag_s\dot{U}_s\sqrt{\rho_0}U^\dag_s-U_s\sqrt{\rho_0}\dot{U}^\dag_sU_s\sqrt{\rho_0}U^\dag_s\notag\\
=&\rho A_U(X)+A_U(X)\rho+\dot{U}_sU^\dag_s\rho-\rho U_s\dot{U}_s^\dag.
\end{align}
From Eq.~(\ref{AUb}) and $U_s\dot{U}_s^\dag=-\dot{U}_sU^\dag_s$, we finally get
\begin{align}\label{upc4}
&U_s\sqrt{\rho_0}U^\dag_s\dot{U}_s\sqrt{\rho_0}U^\dag_s-U_s\sqrt{\rho_0}\dot{U}^\dag_sU_s\sqrt{\rho_0}U^\dag_s\notag\\
=&U_sU_a\dot{U}_a^\dag\rho_0U^\dag_s-U_s\rho_0 \dot{U}_a U_a^\dag U^\dag_s,
\end{align}
which validates Eq.~(\ref{upc}).


\bibliographystyle{apsrev}

\end{document}